\documentclass[onecolumn,aps,prb,showpacs,superscriptaddress]{revtex4}
\usepackage{color}
\usepackage{tabularx}
\usepackage{amsmath}
\usepackage{graphicx}
\usepackage{amssymb}
\usepackage{bm}

\newcommand{\be}{\begin{equation}}
\newcommand{\ee}{\end{equation}}
\newcommand{\ba}{\begin{eqnarray}}
\newcommand{\ea}{\end{eqnarray}}
\begin{document}
\title{Spontaneous polygonization of multi-walled carbon nanotubes: perturbation analysis}
\author{Weihua MU}
\email{muwh@itp.ac.cn} \affiliation{Key Laboratory of Frontiers in
Theoretical Physics and Kavli Institute for Theoretical Physics
China, Institute of Theoretical Physics, The Chinese Academy of
Sciences, P.O.Box 2735 Beijing 100190, China}
\author{Zhong-can Ou-Yang}
\affiliation{Key Laboratory of Frontiers in Theoretical Physics and
Kavli Institute for Theoretical Physics China, Institute of
Theoretical Physics, The Chinese Academy of Sciences, P.O.Box 2735
Beijing 100190, China} \affiliation{Center for Advanced Study,
Tsinghua University, Beijing 100084, China}
\begin{abstract}
Spontaneous polygonization for a multi-walled carbon nanotubes (MWCNTs)
have been observed for about two decades. In present manuscript, 
this phenomenon is understood by the competition between cohesion energy (with lattice mismatching effect) and curvature elastic energy of tubes inside the MWCNTs. Based on a continuum elastic model and perturbation analysis, the crucial conditions for spontaneous polygonization of MWCNT is expressed by fundamental parameters of MWCNTs, which are in reasonable agreement with all relevant 
experiments and computer simulations reported in literatures. Present approach can be used in further studies of crossectional geometry-dependent properties of MWCNTs. 
\end{abstract}
\maketitle
\section{Introduction}
Multi-walled carbon nanotubes (MWCNTs) are concentric cylinders which can be
thought of rolled from graphene layers of graphites. The MWCNT was firstly discovered in 1991~
\cite{iijima91,iijima93}, after that, carbon nanotubes (CNTs) have aroused intense interest due 
to their extraordinary properties and promising potential for nanotechnology 
applications~\cite{saito98}. 

Because of CNTs' special atomic structures, they are expected to exhibit physical 
behaviors fundamentally different from those of graphite~\cite{saito98}. Electronically, a CNT can be either a semiconductor or a metal, depending on its diameter and chirality, and 
the electronic band structure is sensitive to the cross-sectional geometry of the tube. Thus the shape transition of tube cross section may play an important role in the applications of nano-electronic devices and nano-electromechanical system (NEMS)~\cite{tang00,tang02}. 

Some circular crosssectional tubes in a MWCNT can be polygonized in response to hydrostatic pressure~\cite{tang00,tang02}. The pressure induced polygonization is help to form  CNTs bundles with $2$D hexagonal closed-packed structures~\cite{rols01,sluiterl02}. Similar pressure induced polygonization of CNTs has also 
been observed for CNTs under non-hydrostatic pressures~\cite{karmakar03}. Interestingly, 
polygonization can occur spontaneously for MWCNTs with large cross section even without radial pressure~\cite{kiang98,wu05,yoon07}. Kiang {\it et al.}~\cite{kiang98} used x-ray diffraction method, found polygonization (or graphitization) flattens the surface of the tube, decreases the interlayer space, as well as introduces sites with high curvatures. Wu {\it et al.}~\cite{wu05} also studied the structural transition of MWCNTs by x-ray diffraction, suggested the larger the tube cross section is, the easier polygonization occurs. After heat treatment, they found critical diameter for polygonization being $50$ nm (lower bound). Yoon {\it et al.}~\cite{yoon07} also studied the heat treated CNTs, showed that although normal CNTs are stable at low temperature, entropy due to Stone-Wales defects created during heat treatment make polygonal shape of larger CNTs more stable, and the radius of a polygonized tube is $50-100$ nm. All the experiments, also at different conditions, suggested the shape transition of tubes is radius relevant, i.e., there is a critical radius for polygonization of tubes in a MWCNT.  

Spontaneous polygoniztion of tubes in MWCNTs has not been studied extensively by theorists yet, for that atomic simulations are time-consuming and continuum model usually lost some important information on lattice structure. The first continuum model for polygonization is proposed by Golovaty {\it et al.}~\cite{golovaty08}. They considered strong covalent bonds between the atoms in a graphene layer and weak bonds between the atoms in adjacent layers in their continuum theory of MWCNTs. Golovaty {\it et al.}'s work is based on the basic idea: the curvature-induced mismatching between the lattices of the adjacent graphene sheets can be reduced by flattening the walls at the expense of creating the line defects that run parallel to the axis of the tube. Polygonization results from the competition between such two types of interactions, which is controlled by the diameter but not the number of walls in MWCNTs. Tibbetts {\it et al.}~\cite{tibbetts09} proposed a model considering many energy terms, such as bending strain energy, tensile strain energy, interfacial energy {\it etc.}, showed that for a MWCNT with certain number of walls, optimal polygonal shape of CNT depends on the interlayer spacing and the chirality of the tubes.   

Although the physics of spontaneous polygonization is clear, all previous theoretical studies did not explicitly express the critical radius of this shape transition by basic physical parameters of CNTs. Golovaty {\it et al}'s model and geometric approach are beautiful but highly abstract, impossible to give quantitative results, which limits the application of their theory. Tibbetts {\it et al.} modeled the quasi-polygonal crosssectional CNTs as several flattened walls connected by small curved junctions at the corner of polygons, therefore their results rely on too many assumptions on the exact shape of the tube~\cite{tibbetts09}.  
   
\begin{figure}[ht]
\scalebox{0.3}{\includegraphics{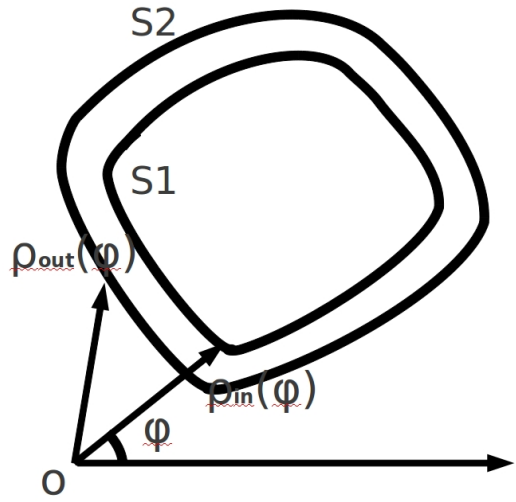}} \caption{\label{fig1} Cross sections of the simplest two-walled CNT.}
\end{figure}

In present manuscript, we will give the explicit expression of the critical radius for polygonization using our geometric theory of continuum elastic model. To use simple model illustrating basic physics, we start with a model containing elastic energy 
(curvature energy) and cohesion energy between adjacent layers,%
\label{energy}
\be
E = E_{cur} + E_{coh}. 
\ee
For simplicity, we consider a straight MWCNTs in which all tube cross sections are perpendicular to the axis. Polygonization is projected to a 2D planar geometric problem, as shown in Fig.~\ref{fig1}.  
A boundary of tube cross section is a closed plane curve described by the 
polar vector $\rho=\rho(\phi)$, with $\phi\in[0,\,2\pi)$ being polar angle. 

Following the approach in Ref.~\onlinecite{golovaty08}, we study the simplest case of two-walled CNTs, since polygonization is sensitive to the radius of tube, but not the number of walls. We assume that cross sections of these two tubes are concentric, similar to each other in the meaning of Euclidean plane geometry. Given the 
the closed curve for inner tube $C_1$ being $\rho_{in}=\rho(\phi)$, obviously, $\rho_{out}=k\,\rho(\phi)$ is the 
radius of the larger tube, and $k$ is the geometric similarity ratio. The ratio for areas enclosed by $C_2$ and $C_1$ is $S_2/S_1=k^2$. 

Only curvature elastic energy is considered in present model, since the the strain energy corresponding to stretching of carbon-carbon bonds are much higher than curvature energy related to rotating the bonds, i.e., changing the bond angles. Curvature energy of a CNT tube have a form of~\cite{yakobson96,oy97}, 
\be\label{membraneenergy}
E_{c}=\frac{k_{c}}{2}\oint(2H)^{2}\, dA+\bar{k}\oint K\, dA,
\ee
Here $H$ and $K$ are the mean curvature and Gaussian curvature for the
tube surface. The bending constants $k_{c}=1.17\,$eV and $\bar{k}/k_c=-1.56$ was obtained by Ou-Yang {\it et al.}~\cite{oy97}, which can be expressed by basic bond parameters of Lenosky type lattice energy ~\cite{lenosky92}, and in good agreement with the values reported by Tersoff {\it et al.}~\cite{tersoff92}.   
The Eq.~(\ref{membraneenergy}) can be projected to 2D elastic energy, 
\be
E^{cuv}=\frac{k_{c}L}{2}\oint k_{r}^{2}\, \mathrm{d}s.
\ee
Here, $k_{r}$ is curvature of plane curve $\rho(\phi)$ as shown in Fig.~\ref{fig1}, $s$ is arc 
parameter of the curve $C_1$, $L$ is tube length.

In general, interlayer interaction can be obtained by the summation of inter-atomic potential between carbon atoms in adjacent layers~\cite{golovaty08}. For simplicity, we use the continuum form of interlayer interaction energy, which is proportional to the volume between inner and outer tubes~\cite{oy97},
\be
\label{EV}
E_{coh}=-g_{0}L\left(k^{2}-1\right)\oint\,\mathrm{dA},
\ee
Here, $\mathrm{d}A$ is area element of area enclosed by curve $S_1$. The $-g_0=\eta\Delta E_c/d_0$, $\Delta E_c=-330\,\mathrm{erg/cm^2}=-2.04\,\mathrm{eV/nm^2}$ is the interlayer cohesion energy of planar graphite obtained theoretically by Girifalco and Lad~\cite{girifalco56,oy97}, $d_0\approx 0.34\,\mathrm{nm}$ is interlayer space of tubes in a MWCNT, which varies little with the increasing of number of shells~\cite{kiang98,wu05,yoon07,tibbetts09}, thus we keep it being a constant. Two adjacent tubes may have different chiralities, leading to incommensurate lattice structures. Typically, the interlayer interaction between two mismatched lattices is smaller than that for the commensurate ones. Therefore, the factor $\eta$ is introduced to account for the mismatching effect between the interlayer lattices, $0<\eta<1$, and $\eta=1$ corresponds to the perfect matched lattices, i.e., graphite case. Exact value of mismatching factor $\eta$ should be determined numerically by first principle calculations. The total energy gives,

\be
\label{totalenergy}
E=E_{cuv}+E_{coh}=\alpha\oint k_{r}^{2}\, ds\,+\,\beta\oint\,\mathrm{d}A,
\ee
Here, constants $\alpha$, $\beta$ and geometric similarity ratio can be expressed as
\[
\alpha =\frac{k_{c}L(k+1)}{2k},\quad \beta =-g_{0}L\left(k^{2}-1\right),\quad k=1+\frac{d_0}{R_{in}}.
\]

Equilibrium shape of the tube's cross section is derived from variation equation $\delta \mathcal{F}=0$, 
and $\mathcal{F}=E+\lambda\left( \oint \mathrm{d}s-L_0\right)$. The perimeters of $C_1$ are fixed due to the carbon-carbon bond lengths are unchanged during spontaneous polygonization. The restriction enters objective function of variation $\mathcal{F}$ by the term with Lagrange factor $\lambda$. Vanishing of $\delta F$ gives the shape equation,
\be
\label{shapeequation}
\alpha k_r^{3}(s)+2\,\alpha\,k_r''(s)+\beta-2\lambda k_r(s)=0,
\ee
where, $"k_r''(s)"$ denotes $\mathrm{d}^2k_r/\,\mathrm{d}s^2$. 

Obviously, circle is one of the solutions of Eq.~(\ref{shapeequation}) with $k_r=1/R_{in}$, since this cubic algebraic equaiton of $1/R_{in}$ always has real roots.   

Polygonization means circular crosssectional tubes are instability, which we will discussed in detail by perturbation analysis. We consider a slight distortion of a circular cross-sectional tube pair with $\rho(\phi)=R_{in}$~\cite{iwamoto04},  

\be\label{perturbation}
\rho(\phi)=R_{in}+\sum_{m}b_{m}e^{im\phi},
\ee
where $\phi\in[0,2\pi),\,m=0,\pm 1,\pm 2,\dots,\pm\infty$, and $b_{m}^{*}=b_{-m}$ ensures $\rho$ is real. High-order harmonic modes are presented in the expression of $\rho(\phi)$ as perturbation terms. The variations of the area enclosed by the curve $C_1$, length of circumference of $C_1$, and the curvature-elastic energy can be obtained as~\cite{oy97},  
 
\ba
\delta\int\mathrm{d}A & = & \pi R_{in}^{2}\left[2\left(b_{0}/R_{in}\right)+\sum_{m=0}^{\infty}\left|\frac{b_{m}}{R_{in}}\right|^{2}\right],\label{deltaA}\\
\delta\int ds & = & \pi R_{in}\left[2\left(\frac{b_{0}}{R_{in}}\right)+\sum_{m=0}^{\infty}m^{2}\left|\frac{b_{m}}{R_{in}}\right|^{2}\right],\label{deltas}\\
\delta\int k_{r}^{2}\, ds & = & \pi R_{in}^{-1}\left[-2\left(\frac{b_{0}}{R_{in}}\right)+\sum_{m=0}^{\infty}\left(2m^{4}-5m^{2}+2\right)\left|\frac{b_{m}}{R_{in}}\right|^{2}\right].\label{deltak2}
\ea

The length of $C_1$ is conserved, therefore, $\delta\int\,\mathrm{d}s=0$, which provides one restriction on coefficient $b_0$,
\be\label{restriction}
\frac{b_{0}}{R_{in}}=-\frac{1}{2}\sum_{m=0}^{\infty}m^{2}\left|\frac{b_{m}}{R_{in}}\right|^{2}.
\ee

Substitute the restriction Eq. (\ref{restriction}) to Eq.(~\ref{deltaA}) and Eq.~(\ref{deltak2}), and the total energy shown as Eq.~(\ref{totalenergy}), gets, 
\ba
\delta\int\mathrm{d}A & = & \pi R_{in}^{2}\left[\sum_{m=0}^{\infty}\left(1-m^{2}\right)\left|\frac{b_{m}}{R_{in}}\right|^{2}\right],\\
\delta\int\, k_{r}^{2}\, ds & = & 2\pi R_{in}^{-1}\left[\sum_{m=0}^{\infty}\left(m^{2}-1\right)^{2}\left|\frac{b_{m}}{R_{in}}\right|^{2}\right].
\ea
yields the deformation energy, 
\ba\label{deltaE}
\delta E & \sim & \pi R_{in}^{-1}\sum_{m=0}^{\infty}\left[2\alpha\left(m^{2}-1\right)-\beta R_{in}^{3}\right]\left(m^{2}-1\right)\left|\frac{b_{m}}{R_{in}}\right|^{2}.\nonumber
\ea

The trivial case of $m=1$, characterized by $\delta E=0$, means a translation of the circle. The $m$-th harmonic deformation can happen when the coefficient of $|b_m/R_{in}|^2$ in Eq.~(\ref{deltaE}) becomes negative, which is, 
$R_{in}>R_{m}$, with $R_m\equiv \left[ 2\alpha (m^2-1)/\beta \right]^{1/3},\,(m\geq 2)$. The occurring of $m$-order harmonic mode lowers the energy, as well as lowers the symmetry of tube from $C_{\infty}$ to $C_m$. Substitute the expressions of $\alpha$ and $\beta$ to $R_{in}>R_m$ gives,
\be\label{inequ}
R_{in}>\sqrt\frac{k_c}{g_0\,d_0}\times\sqrt{m^2-1},
\ee

Experiments suggests hexagon is a typical quasi-polygonal shape of distorted CNTs, 
thus the critical radius for the occurring of hexagonal cross section is 
\be\label{Rcritical}
R_{critical}\approx 6\,\sqrt\frac{k_c}{\eta\Delta E_c},
\ee 
Tubes with radius larger than $R_{critical}\approx 5/\sqrt{\eta}\,\mathrm{nm}$ should be polygonized to hexagonal shape. For $\eta=0.1$, $R_{critical}=15\,\mathrm{nm}$, which is in reasonable accord with the spontaneous polygonization phenomena of MWCNTs observed in recent experiments~\cite{kiang98,yoon07}. Based our calculation, we confirme that mismatching of lattices (reflected by factor $\eta<1$) plays an important role in polygonization, and mismatching effect can be amplified by the defects introduced in the process of heat treatment.  

In summary, we have studied the spontaneous polygonization of CNTs without applied hydrostatic pressure. The key step is the perturbation analysis for the instability of original circular crosssectional tube with large circumference. We only consider the bending energy coming from the interatomic interaction for atoms in the same layer and cohension
energy due to interatomic interaction for atoms in adjacent layers, as well as the mismatching of lattices. With the length of circumference being fixed, the 
competition of bending energy and cohension energy lends to the polygonization. We obtained the the critical radius for tube occurring shape transition from circular shape to hexagonal shape, as shown in Eq.~(\ref{Rcritical}), which is the first quantitatively expression for the critical radius of spontaneous polygonization, The results are in accord with recent experiments, and can be used to predict polygonization related phenomena. 

The work is supported by National Science Foundation of China (NSFC) under Grants No. 11074259.

\newpage
\newpage
\begin{figure}[ht]
\scalebox{0.3}{\includegraphics{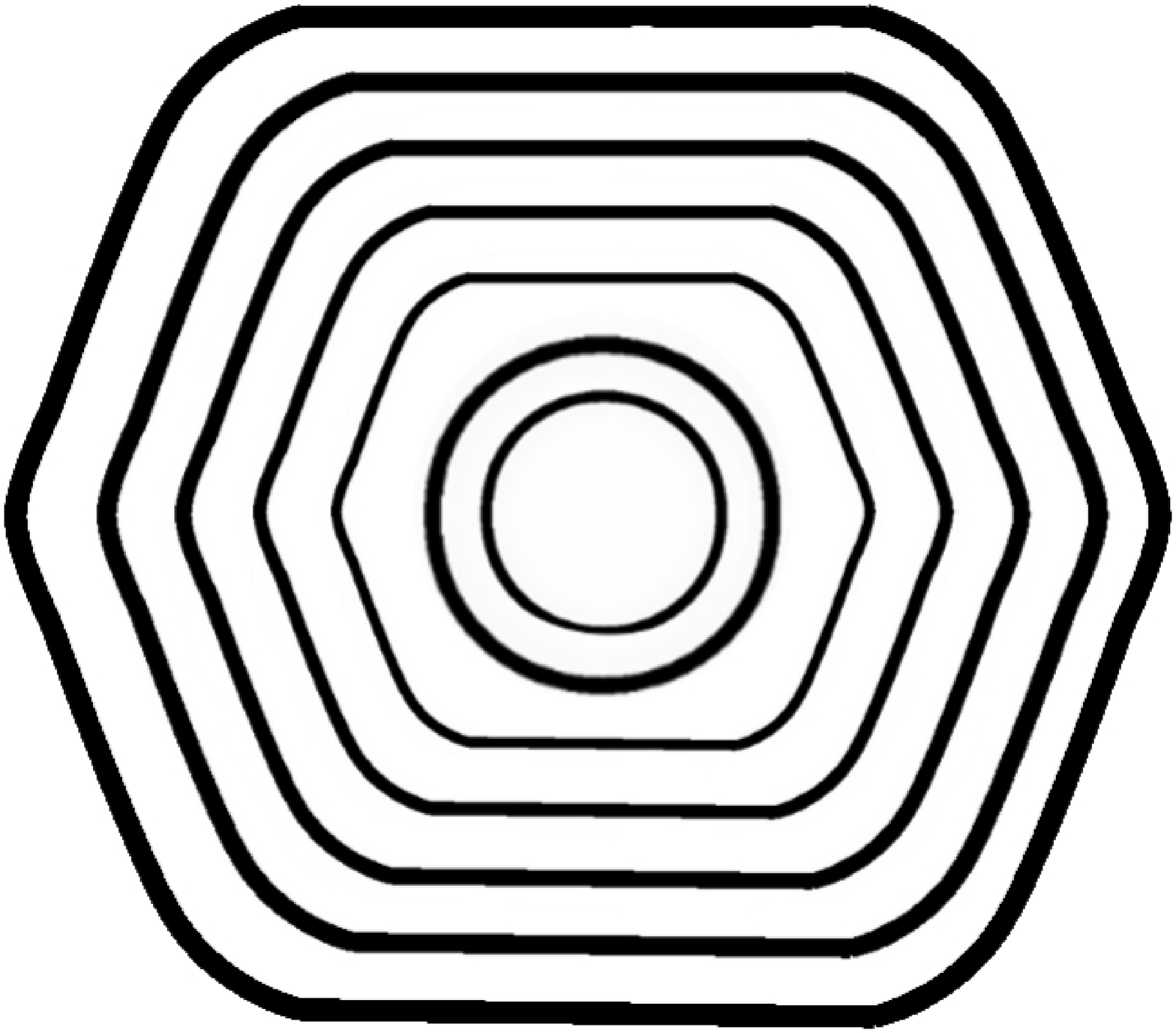}} \caption{\label{fig2} Spontaneous polygonization for the tubes with larger diameter in a MWCNT.}
\end{figure}


\begin{thebibliography}{99}
\bibitem{iijima91}
S. Iijima, Nature (London), {\bf 354}, 56~(1991). 

\bibitem{iijima93}
S. Iijima and T. Ichihashi, Nature, {\bf 363}, 603~(1993).

\bibitem{saito98}
R. Saito, M. S. Dresselhaus, G. Dresselhaus, {\textit Physical Properties of Carbon Nan-
otubes}, Imperial College Press, (London, 1998).

\bibitem{tang00}
J. Tang, L-C. Qin, T. Sasaki, M. Yudasaka,
A. Matsushita, and S. Iijima, Phys. Rev. Lett. {\bf 85}, 1887~(2000).

\bibitem{tang02}
J. Tang, L-C. Qin, T. Sasaki1, M. Yudasaka,
A. Matsushita1 and S. Iijima, J. Phys.: Condens. Matter {\bf 14}, 10578~(2002).

\bibitem{rols01}
S. Rols, I. N. Goncharenko, R. Almairac, J.
L. Sauvajol, and I. Mirebeau, Phys. Rev. B {\bf 64}, 153401~(2001).

\bibitem{sluiterl02}
M. H. F. Sluiter1, V. Kumar and
Y. Kawazoe1, Phys. Rev. B {\bf 65}, 161402~(2002).

\bibitem{karmakar03}
S. Karmakar, S. M. Sharma, P. V. Teredesai,
D. V. S. Muthu, A. Govindaraj, S. K. Sikka, and A. K. Sood, New. J. Phys. {\bf 5} 143~(2003).

\bibitem{kiang98}
C. -H. Kiang, M. Endo, P. M. Ajayan, G. Dresselhaus,
and M. S. Dresselhaus, Phys. Rev. Lett. {\bf 81}, 1869~(1998).

\bibitem{wu05}
F. Y. Wu, and H. M. Cheng, J. Phys. D {\bf 38}, 4302~(2005).

\bibitem{yoon07}
M. Yoo, J. Howe, G. Tibbetts, G. Eres, and Z. Zhang, Phys. Rev. B {\bf 75}, 165402~(2007).

\bibitem{golovaty08}
D. Golovaty, and S. Talbott, Phys. Rev. B {\bf 77}, 081406(R)~(2008).

\bibitem{tibbetts09}
K. Tibbetts, R. Doe, and G. Ceder, Phys. Rev. B {\bf 80}, 014102~(2009).

\bibitem{yakobson96}
B. I. Yakobson. C. J. Brabec; J. Bernholc, Phys. Rev. Lett. {\bf 76}, 2511~(1996).

\bibitem{oy97}
Ou-Yang Z-c, Z-B. Su, and C-L. Wang, Phys. Rev. Lett. {\bf 78}, 4055~(1997).

\bibitem{lenosky92}
T. Lenosky, {\it et al.} Nature {\bf 355}, 333~(1992).
\bibitem{tersoff92}
J. Tersoff, Phys. Rev. B {\bf 46}, 15546~(1992).

\bibitem{girifalco56}
L. A. Girigalco, and R. A. Lad, J. Chem. Phys. {\bf 25}, 693~(1956).

\bibitem{iwamoto04}
M. Iwamoto, and Ou-Yang zhong-can, Phys. Rev. Lett. {\bf 93}, 206101~(2004).
\end{thebibliography}
\end{document}